\documentclass{jfm}

\usepackage{graphicx}
\usepackage{epstopdf, epsfig}
\usepackage{graphicx}
\usepackage{dcolumn}
\usepackage{bm}
\usepackage{color}
\usepackage{array}
\usepackage{amsmath,amssymb,stmaryrd} 
\usepackage{multicol}

\usepackage[abs]{overpic}
\usepackage{subcaption}

\definecolor{blue}{rgb}{0, 0.4470, 0.7410}
\definecolor{red}{rgb}{0.8500, 0.1250, 0.0480} 
\definecolor{orange}{rgb}{0.8500, 0.3250, 0.0980} 
\definecolor{yellow}{rgb}{0.9290, 0.6940, 0.1250}
\definecolor{purple}{rgb}{0.4940, 0.1840, 0.5560}
\definecolor{green}{rgb}{0.4660, 0.6740, 0.1880}
\definecolor{ltblue}{rgb}{0.3010, 0.7450, 0.9330}
\definecolor{dkred}{rgb}{0.6350, 0.0780, 0.1840}
\definecolor{gray}{rgb}{0.22, 0.22, 0.3}

\shortauthor{Q. Liu, B. An, M. Nohmi, M. Obuchi, \& K. Taira} 
\title{Core-pressure alleviation for a wall-normal vortex by active flow control }

\author{Qiong Liu\aff{1},
Byungjin An\aff{2},
Motohiko Nohmi\aff{2},\\
Masashi Obuchi\aff{2}, \and
Kunihiko Taira\aff{1}
\corresp{\email{ktaira@fsu.edu}} 
}

\affiliation{\aff{1} Department of Mechanical Engineering, Florida State University, Tallahassee, FL 32310, USA
\aff{2} Advanced Analysis Department, Ebara Corporation, Tokyo, 144-8510, Japan
}

\begin{document}

\maketitle

\begin{abstract}
We consider the application of active flow control to modify the radial pressure distribution of a single-phase wall-normal vortex.  The present flow is based on the Burgers vortex model but with a no-slip boundary condition prescribed along its symmetry plane.  The wall-normal vortex serves as a model for vortices that emerge upstream of turbomachinaries, such as pumps.  This study characterizes the baseline vortex unsteadiness through numerical simulation and dynamic mode decomposition.  The insights gained from the baseline flow are used to develop an active flow control technique with rotating zero-net-mass blowing and suction for the objective of modifying the core pressure distribution.  The effectiveness of the control strategy is demonstrated by achieving a widened vortex core with increased pressure.  Such change in the flow field weakens the local strength of the wall-normal vortex core, potentially inhibiting the formation of hollow-core vortices, commonly encountered in liquids.
\end{abstract}

\begin{keywords}
vortex dynamics, flow control, stability.
\end{keywords}

\section{Introduction}

The appearance of vortical structures in fluid-based machines is often associated with the reduction in their efficiency and performance.  The generation of large wake vortices, contributes to increased drag on bodies and the presence of near-wall vortices in turbulent flows, produces higher wall shear stress compared to that of laminar flows for both internal and external flows. Added complications can arise when these vortices emerge in liquids. Low-pressure cores of strong vortices can become hollow due to cavitation or accumulation of gas bubbles present in the flow.  When such hollow-core vortices are engulfed by pumps, not only degrade the operating performance but they also can cause unsteady loading, potentially damaging the structures \citep{Zhao:Vis10, Brennen11}.  In many fluid-based applications, sustained operation of these machines at desired levels of performance is of critical importance, especially when they are utilized for chemical production, cooling in nuclear power plants, and storm surge response \citep{An:Ebara18}. 

To date, there have been numerous studies on implementing passive flow control to prevent the vortex generation or to alleviate the core-pressure effects in turbomachineries, including anti-vortex baffles \citep{yang2017computational} and geometric modifications. Nonetheless, passive control does not offer the ability to adjust the control efforts adaptively. The design of efficient and reliable active flow control techniques for the vortical flow associated with pumps is still under explored, but has great potential to improve their sustained operations in critical applications. 

\begin{figure}
\centering
   \includegraphics[width=0.85\textwidth]{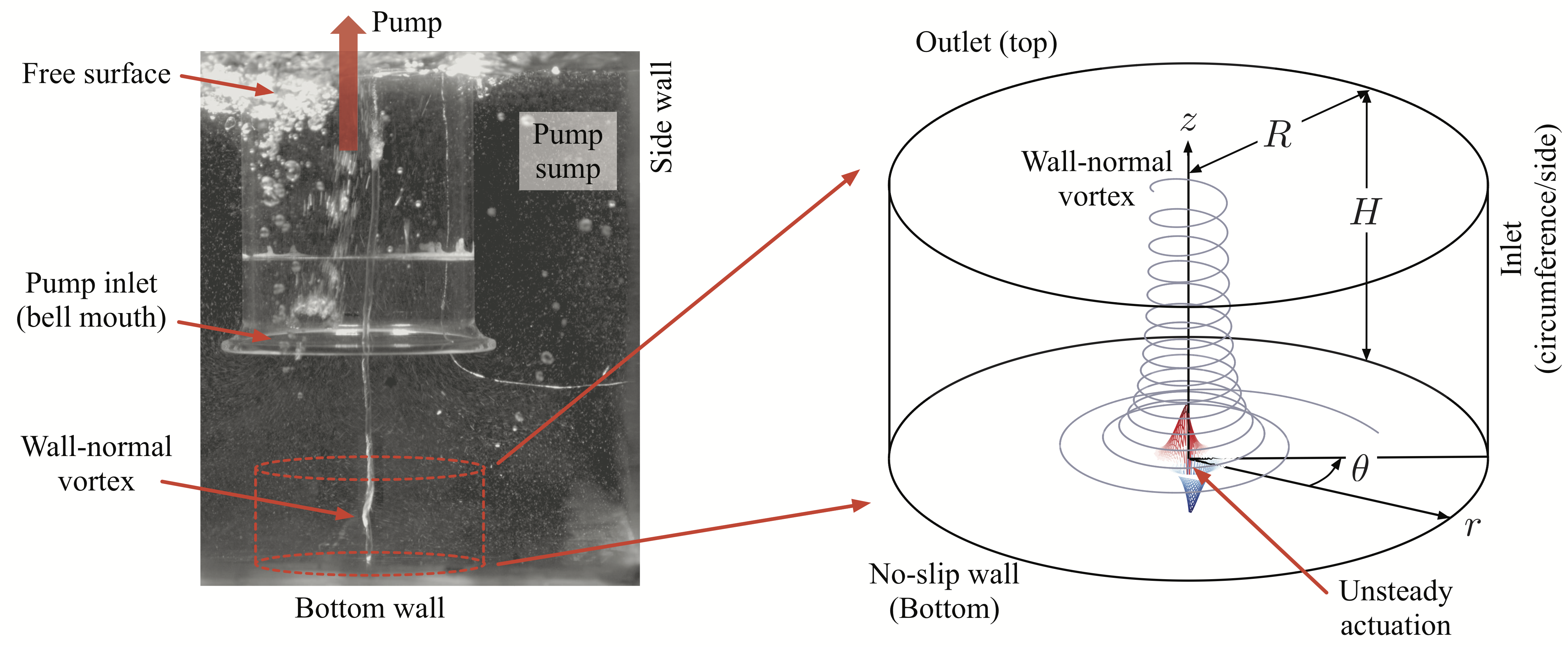}
   \caption{Left: experimental visualization of a submerged wall-normal vortex generated under an inlet in a pump sump (water).  A second wall-normal vortex terminating at the side wall can also be observed.  Right: the computational setup used in the present flow control study to analyze the model vortex.}
   \label{fig:setup}
\end{figure}

In the present study, we consider a wall-normal vortex that emerges upstream of turbomachineries.  An example of such a hollow-core vortex in water is exhibited in figure \ref{fig:setup} (left), highlighted within the box (red dash line). We model this type of vortex as a single-phase wall-normal vortex in the computational study, as illustrated in figure \ref{fig:setup} (right). The flow enters the computational domain from the circumferential side over a no-slip wall at the bottom and exits through the top. The vortical flow is simulated by prescribing the Burgers vortex \citep{Burgers:AAM48} velocity profile along the inlet plane.  

The Burgers vortex is the exact steady solution to the incompressible Navier--Stokes equations \citep{Burgers:AAM48}, which has been studied in detail in the past \citep{Saffman92,Wu2006}.  In particular, its stability properties have been analyzed by \cite{Leibovich:PF81}, \cite{Crowdy:SAM98}, and \cite{Schmid:JFM04}.  The linear three-dimensional stability analysis of \cite{Schmid:JFM04} reports that Burgers vortex is neutrally stable but can exhibit transient energy growth.  Such behavior can alter the mean flow field in full nonlinear flow. In the present study, we aim to introduce forcing inputs that leverage the transient growth to alter the radial profile of the vortex core for pressure increase to alleviate the aforementioned adverse effects.

In what follows, we present our approach to model the single-phase wall-normal vortex in section \ref{sec:approach} and describe its baseline characteristics in section \ref{sec:baseline}.  Dynamic mode decomposition is performed on the baseline flow field to gain insights into the oscillatory dynamics of the wall-normal vortex and to develop an effective flow control technique.  In section \ref{sec:control}, we apply active flow control based on the baseline characterization to effectively modify the radial profile of the vortex and increase the core pressure.  At last, the paper offers some concluding remarks in section \ref{sec:conclusion}.  

\section{Problem description and approach}
\label{sec:approach}

We numerically study the single-phase wall-normal vortex and the influence of unsteady actuation on its core profile. A cylindrical computational domain is considered. The inflow enters from the circumferential side and leaves from the top boundary, where a convective outflow condition is provided.  Along the bottom wall, the no-slip wall boundary condition is prescribed. The swirling flow is introduced from the circumferential inflow boundary using the laminar velocity profile of the Burgers vortex \citep{Burgers:AAM48}
\begin{equation}
   u_r=-\frac{1}{2}\gamma r, \quad 
   u_\theta=\frac{\Gamma_\infty}{2\pi r}\left[1-\exp\left(-\frac{r^2}{a^2}\right)\right], \quad 
   u_z=\gamma z
\end{equation}
at $r = R$ and $0 \le z \le H$, where $\Gamma_\infty$ is the circulation of the Burgers vortex model and $\gamma$ is the strain rate imposed on the flow.  The Burgers vortex has a maximum rotational velocity $u_{\theta,\max}$ at $r = 1.12a$, where $a$ is the core radius. In the present work, we report our results scaled by $a$ for length, $u_{\theta,\max}$ for velocity, and $p^\ast=\frac{1}{2}\rho u_{\theta,\max}^2$ for pressure, where $\rho$ is the density of the fluid.  The non-dimensional parameter relevant for this study is the circulation-based Reynolds number, which is set to $Re \equiv \Gamma_\infty/\nu = 5000$ with $\nu$ being the kinematic viscosity.   The strain rate is determined from the choice of Reynolds number and the core radius $a$ such that $\gamma = 4\nu/a^2$.  The circulation based on the present scaling is $\Gamma_\infty a/u_{\theta,\max} = 9.848$.

Direct numerical simulations are performed using a second-order finite-volume based solver, Cliff, to numerically solve the incompressible Navier--Stokes equations \citep{Ham:CTR04}.  The size of the domain is chosen to be $R/a = H/a = 15$.  We have compared results from this domain with those from a larger domain of $R/a = H/a = 20$ to ensure that the domain size is sufficiently large. The maximum difference in the local pressure value is limited to 3\%. The present computational domain is discretized using a structured mesh with 8.4 million grid points, which was deemed adequate in comparison to results based on a mesh with 15.6 million grid points, which showed only a maximum local pressure difference of 0.6\%.  The actuator setup is described later in section \ref{sec:ctrl_setup}.

\section{Baseline Flow}
\label{sec:baseline}

\begin{figure}
   \centering
   \includegraphics[width=0.99\textwidth]{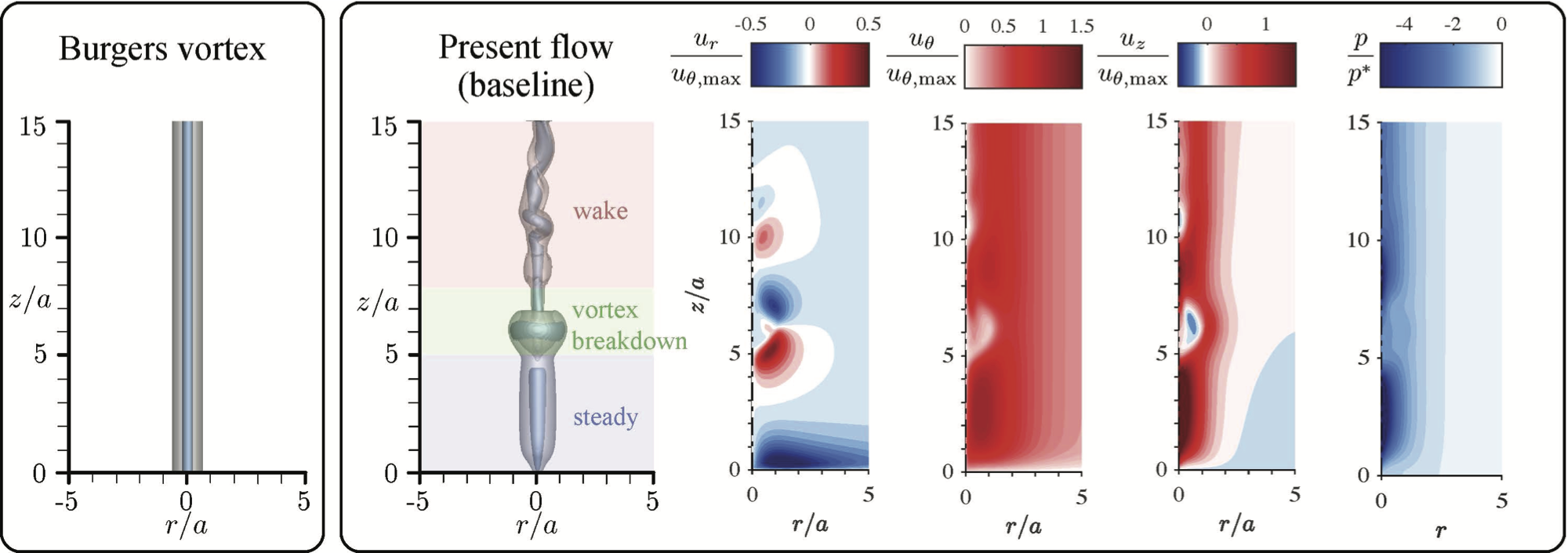}
   \caption{Instantaneous visualizations of the Burgers vortex model and the present baseline vortex.  The two left plots show the vortical structures based on $|| \boldsymbol{\omega} ||=2$ (gray) and $Q$-criterion $Q=2$ (blue).  Steady (blue), vortex breakdown (green), and wake (red) regions are highlighted for the present flow.  Time-averaged velocity and pressure distributions are also shown for the present flow.}
   \label{fig:baseline}
\end{figure}

We examine the baseline wall-normal vortex generated by imposing the Burgers vortex inlet velocity profile.  The current vortical flow differs from the Burgers vortex model since a laminar boundary layer develops along the bottom wall and the resulting vorticity flux is engulfed into the wall-normal vortex.  Visualized in figure \ref{fig:baseline} are the vortex cores of the Burgers vortex and the current vortical flow.  While the Burgers vortex is a steady solution to the Navier--Stokes equations with a constant core radius along the axial ($z$) direction, the current flow model exhibits three different regions of the flow characterized as (i) steady: $ 0 < z/a \lesssim 5$; (ii) vortex breakdown: $5 \lesssim z/a \lesssim 8$; and (iii) wake: $8 \lesssim z/a$, as highlighted in color in figure \ref{fig:baseline}. 

In the steady region ($0 < z/a \lesssim 5$), the vortex structure remains axisymmetric as shown in figure \ref{fig:baseline}.  The swirling flow forms a cone-shaped vortex above the wall due to the wall effect as shown by the velocity profiles in figure \ref{fig:baseline}.   As the flow convects upward from the sides, the core becomes uniform and resembles the Burgers vortex ($z/a \lesssim 5$).  Inward radial flow, quantified by radial velocity $u_r$, emerges adjacent to the wall until it reaches near the center $(x/a, z/a)=(0,0)$.  The inward radial flow then redirects upward to provide a strong positive axial velocity $u_z$.  Once the flow turns upward, the azimuthal velocity $u_{\theta}$ reaches its maximum value at $r/a \approx 1.12$ away from the wall. In the steady flow region, a low-pressure region emerges in the vortex core, which is the target region of flow control to achieve increase in the pressure distribution, as we explore later.

The vortex experiences a breakdown for $5 \lesssim z/a \lesssim 8$ in the flow, which is one of the key differences between the current vortical flow and the Burgers vortex.  This breakdown is axisymmetric and bulges outward creating a spherical shape \citep{hall1972vortex}, as highlighted in green in figure \ref{fig:baseline}.  The vorticity generated at the wall advects along the vortex cores and leads to the emergence of the adverse pressure gradient.  This causes the flow to expand radially, in a manner similar to flow separation.  As a result, a recirculating flow is created within the breakdown region with low-amplitude oscillation.  Further details of general vortex breakdown can refer to \cite{brown1990axisymmetric} and \cite{kurosaka2006azimuthal}.  

\begin{figure}
   \centering
   \includegraphics[width=\textwidth]{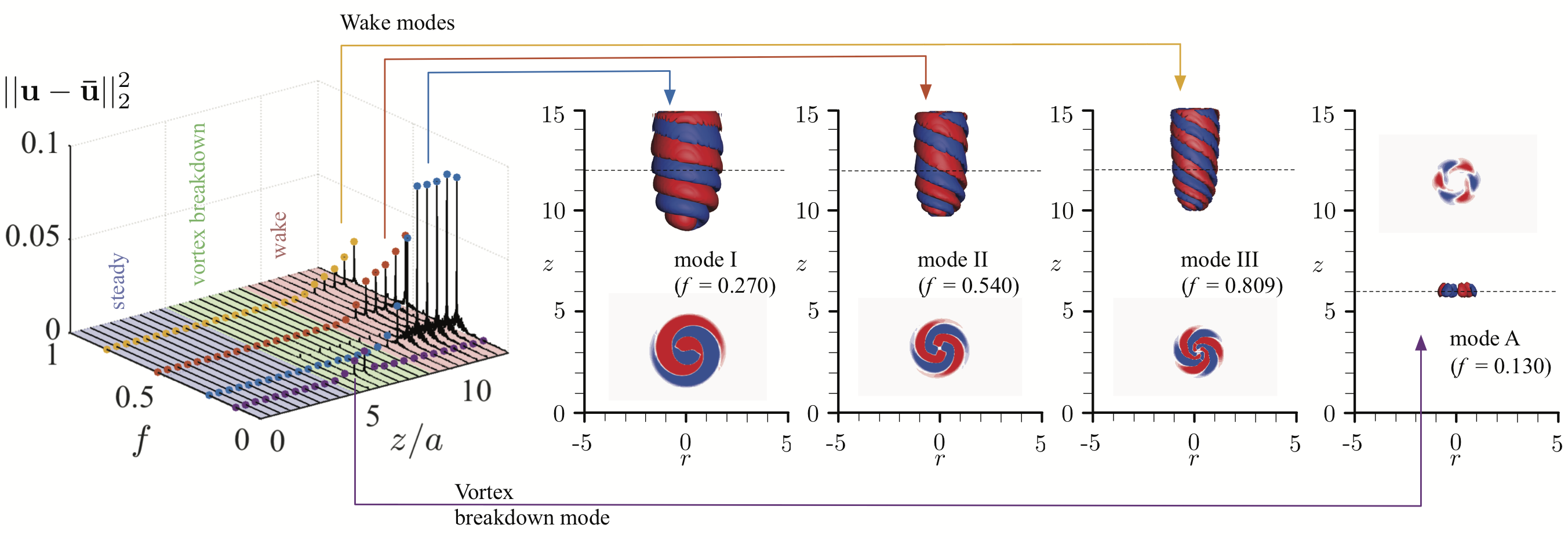}
   \caption{Characterization of the baseline flow unsteadiness.  Kinetic energy spectra along $(r/a, \theta) = (1,0)$ is shown on the left.  DMD modes ($u_z$) corresponding to the spectra peaks are shown on the right (axial slices shown as inserts at the indicated heights).}
   \label{fig:baseline2}
\end{figure}

Downstream of the breakdown region, the flow transitions to the wake region for $8 \lesssim z/a$.  Here, the vortex structure loses its axisymmetry and develops an unsteady spiraling wake.  While the wake region has low pressure, its value is higher than the steady region of the flow. The level of unsteadiness is the highest amongst the three regions of the flow. By analyzing the frequency spectra of the kinetic energy along $(r/a,\theta) = (1,0)$, we observe four dominant frequencies in the flow, as identified in figure \ref{fig:baseline2}.  As mentioned earlier, the vortex breakdown mode (referred to as mode A, hereafter) appears first with a low amplitude but disappears in the wake region.  In the wake region, three modes appear (modes I, II, and III, hereafter) holding higher levels of fluctuations.  

To extract the spatial structures of these modes, we perform modal analysis on the flow field data \citep{Taira:AIAAJ17}.  In particular, we use dynamic mode decomposition (DMD) \citep{Schmid:JFM10, Rowley:JFM09} to extract the spatial modes that correspond to the peaks from the spectral analysis.  DMD eigenvalues obtained from analyzing the snapshots reveal that the fluctuations are sustained (neutrally stable).  The spatial structures of the corresponding dynamic modes are visualized on the right side of figure \ref{fig:baseline2}.  The three dominant wake modes, I, II, and III, exhibit helical structures rotating with the base flow. The azimuthal structures of those three wake modes, shown as inserted two-dimensional plots taken from the planes indicated by the dashed lines, exhibit azimuthal wavenumbers of $m=1$, $2$, and $3$. For this flow, mode I ($m=1$) is found to be the dominant mode. This observation is in agreement with the report by \cite{Schmid:JFM04} that states that, while the Burgers vortex is asymptotically stable, a significant short-term energy amplification emerges for perturbations with azimuthal wavenumber of $m=1$.  Modes II and III are the superharmonics of mode I.  We also identify the vortex breakdown mode A that fluctuates with a lower amplitude.  Mode A only exists in the vortex breakdown region and has a lower oscillation frequency.  This breakdown mode does not possess a strong spiraling structure as seen for the wake modes.

With these modal structures present, the pressure distribution becomes higher than that of the steady region.  As such, we consider perturbing the flow with actuation frequency in the vicinity of the baseline fluctuations to stimulate the emergence of unsteadiness. As presented below, we introduce actuation from the bottom wall to alter the vortex core profile especially in the steady region for increasing the core pressure.


\section{Flow control}
\label{sec:control}
 
\subsection{Control approach}
\label{sec:ctrl_setup}

Based on insights gained from the baseline characterization of the wall-normal vortex, we develop an active flow control strategy to weaken the vortex core. We aim to widen the vortex core radius by inducing unsteadiness in the core region.  Such modification will reduce the azimuthal velocity and hence increase the vortex core pressure.

To trigger the emergence of vortical unsteadiness, we use unsteady actuation from the bottom wall at the vortex center. We have observed from the above DMD analysis that modal structures appear at frequencies between $0.13$ and $0.81$. To introduce unsteady perturbation with these characteristics, we add rotary blowing and suction perturbation by prescribing a velocity boundary condition at the wall:
\begin{equation}
   \frac{u_z^c(r,\theta,z=0,t)}{u_{\theta, \max}}= A_c\cos(2\pi f_c t + m_c\theta ) \exp\left(-\frac{r^2}{a^2}\right),
   \label{eq:actuation}
\end{equation} 
where $A_c$, $f_c$ and $m_c$ are the actuation amplitude, frequency, and azimuthal wavenumber, respectively. Control frequency is chosen from $0.08 \le f_c \le 0.3$ to span across the dominant frequencies identified from the above DMD analysis of the baseline flow. The azimuthal wavenumber is set to $m_c = 1$ (co-rotating) and $-1$ (counter-rotating) as DMD analysis reveals the largest fluctuations at these wave numbers.  Here, there is equal and opposite amplitude of blowing and suction applied to the flow with zero net mass addition. For the sake of completeness, we also consider the actuation with azimuthal wavenumbers $m=0$ and $|m|>1$, as well as steady forcing with $f_c=0$.  This particular form of the function is chosen to implement suction and blowing in a spatially compact manner near the vortex center.  As we briefly discuss later, unsteady actuation with $f_c > 0$ is indeed required to achieve effective control.

To quantify the control input, we introduce a circulation-based oscillatory momentum coefficient defined as 
\begin{equation}
   C^\Gamma_\mu \equiv \frac{J'}{\Gamma_\infty^2} ,
   \quad {\text{where}} \quad
   J' \equiv
   f_c \int_0^{\frac{1}{f_c}} \int_0^R \int_0^{2\pi} 
   (u_z^c)^2 {\rm d}\theta {\rm d}r {\rm d}t 
\end{equation}
is the oscillatory momentum input. We choose control amplitude from $0.1\le A_c \le 1$, which corresponds to $0.02\% \le C^\Gamma_\mu \le 2.03\%$ to evaluate the momentum coefficient effect.

\subsection{Controlled flow}

\begin{figure}
\centering
\includegraphics[width=0.85\textwidth]{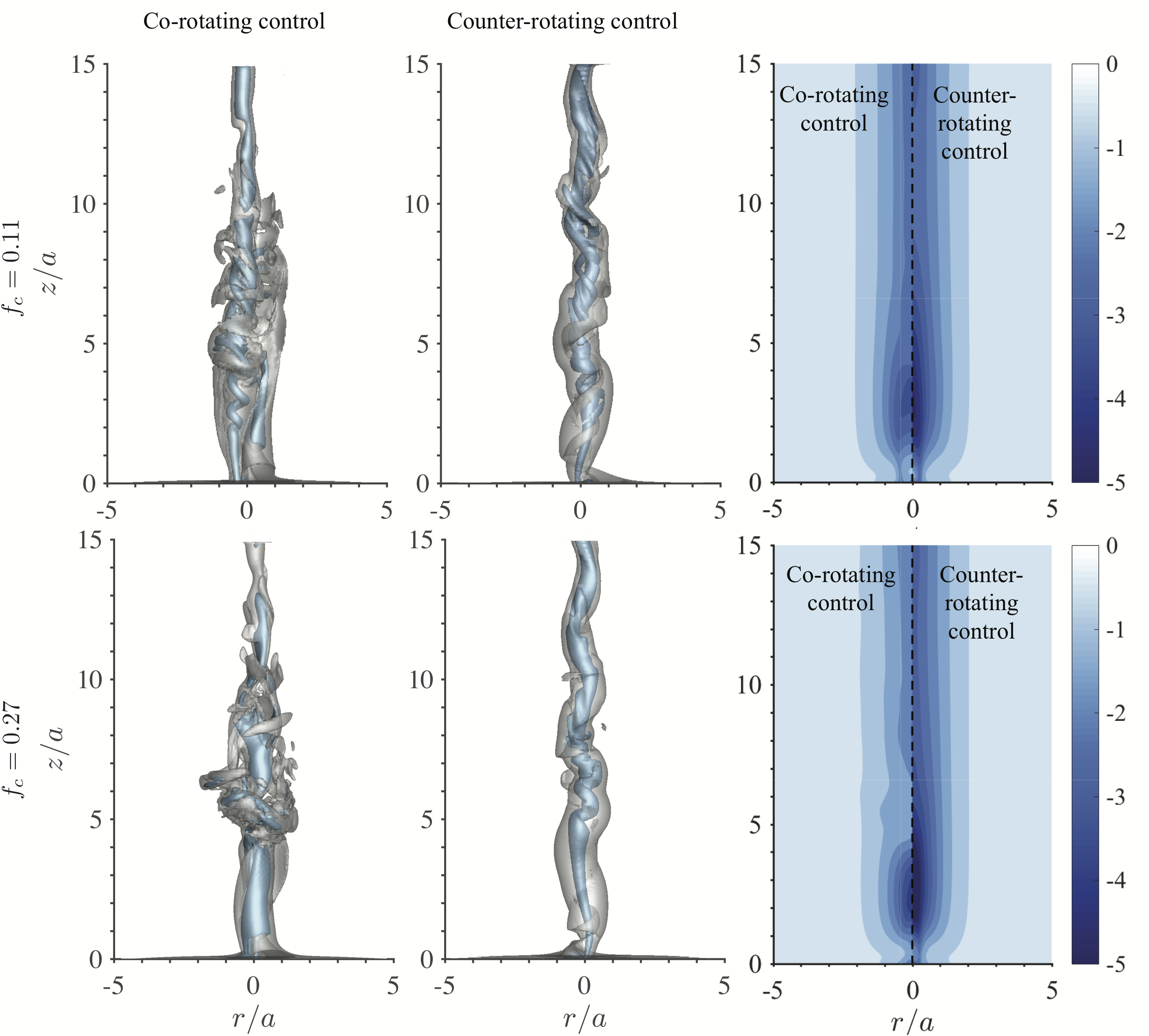}
\caption{Visualizations of the controlled flows with isosurfaces of $||\boldsymbol{\omega}||=2$ (grey) and $Q$-criterion $Q=2$ (blue).  Shown are cases with co-rotating (left column) and counter-rotating (middle column) control using $C_\mu^\Gamma = 2.03\%$ at $f_c = 0.11$ (top) and $0.27$ (bottom).  The corresponding time-averaged pressure contours are also shown (right).}
\label{fig:vortex_strucuture}
\end{figure}

Let us examine the effects of forcing on the wall-normal vortex and its core pressure distribution.  We first set the actuation amplitude at $A_c = 1$ ($C^\Gamma_\mu = 2.03\%$) and study the influence of unsteady forcing at frequencies of $f_c = 0.11$ and $0.27$. These control frequencies are selected to be close to those of baseline DMD modes A and I, with the aim to stimulate the emergence of instabilities around the vortex.  We also consider co-rotating and counter-rotating actuation with $m = \pm 1$.  Presented in figure \ref{fig:vortex_strucuture} are the controlled flows with the chosen control parameters.  Visualized are the vortex cores using the $Q$-criterion isosurface (blue) and the overall vortical structures with the transparent gray vorticity norm isosurface.  The corresponding time-averaged pressure profiles are also shown in figure \ref{fig:vortex_strucuture} to assess the effectiveness in modifying the pressure distribution.

Perturbing the flow at a frequency of $f_c = 0.11$, the vortex profile can be modified significantly with co-rotating and counter-rotating inputs.  As shown in figure \ref{fig:vortex_strucuture} (top), the vortex core is spread radially in an unsteady manner and the vortex breakdown region is eliminated.  In the case of co-rotating control, the vortical structure becomes more complex and radially spread out compared to the baseline flow.  Hence, the core region of the flow has lower azimuthal velocity and leads to an increase in the core pressure, which can be seen in figure \ref{fig:vortex_strucuture} (top-right).  With counter-rotating control, the vortex core is modified, while its width has not been significantly altered. Instead, the core deforms into a braid-like helical structure.  The corresponding minimum pressure in the counter-rotating controlled flow increases by 33\% compared to the baseline flow, but not as much as what is achieved with co-rotating forcing case, which accomplished a 54\% increase.

The wall-normal vortex has also been modified by forcing with a higher frequency of $f_c = 0.27$, as illustrated in figure \ref{fig:vortex_strucuture} (bottom).  With the counter-rotating control, the vortex breakdown region is eliminated and the vortical flow becomes unsteady throughout the entire vortical region. For co-rotationg control, the vortex breakdown is essentially removed but shows slight bulging near $z/a \approx 6$.  For this reason, the pressure increase at the base of the vortex is not achieved as in the lower frequency forcing case.  This suggests that instigating fluctuations at low frequency leads to a high-level of unsteadiness and mixing.  On the other hand, counter-rotating actuation at this higher frequency does not exhibit noteworthy difference from the low frequency actuation.

We further analyze the influence of control frequency and amplitude on the wall-normal vortex.  As discussed above, pressure increase along the vortex core is achieved by spreading the vortex core radially and reducing the azimuthal velocity $u_\theta$.  Hence, we can use the time-averaged axial circulation  
$\Gamma_z(r,z)=\int_0^{2\pi} \bar{u}_\theta(r,z,\theta) r {\rm d}\theta$, normalized by $\Gamma_\infty$ as a metric to assess the effectiveness of the current flow control approach.  {\color{red}}In what follows, we focus on the co-rotating control setup, as it is more effective than the counter-rotating control setup in alleviating the low-pressure core.

\begin{figure}
\begin{tabular}{ll}
    (a) & (b) \\
    \includegraphics[width=0.47\textwidth]{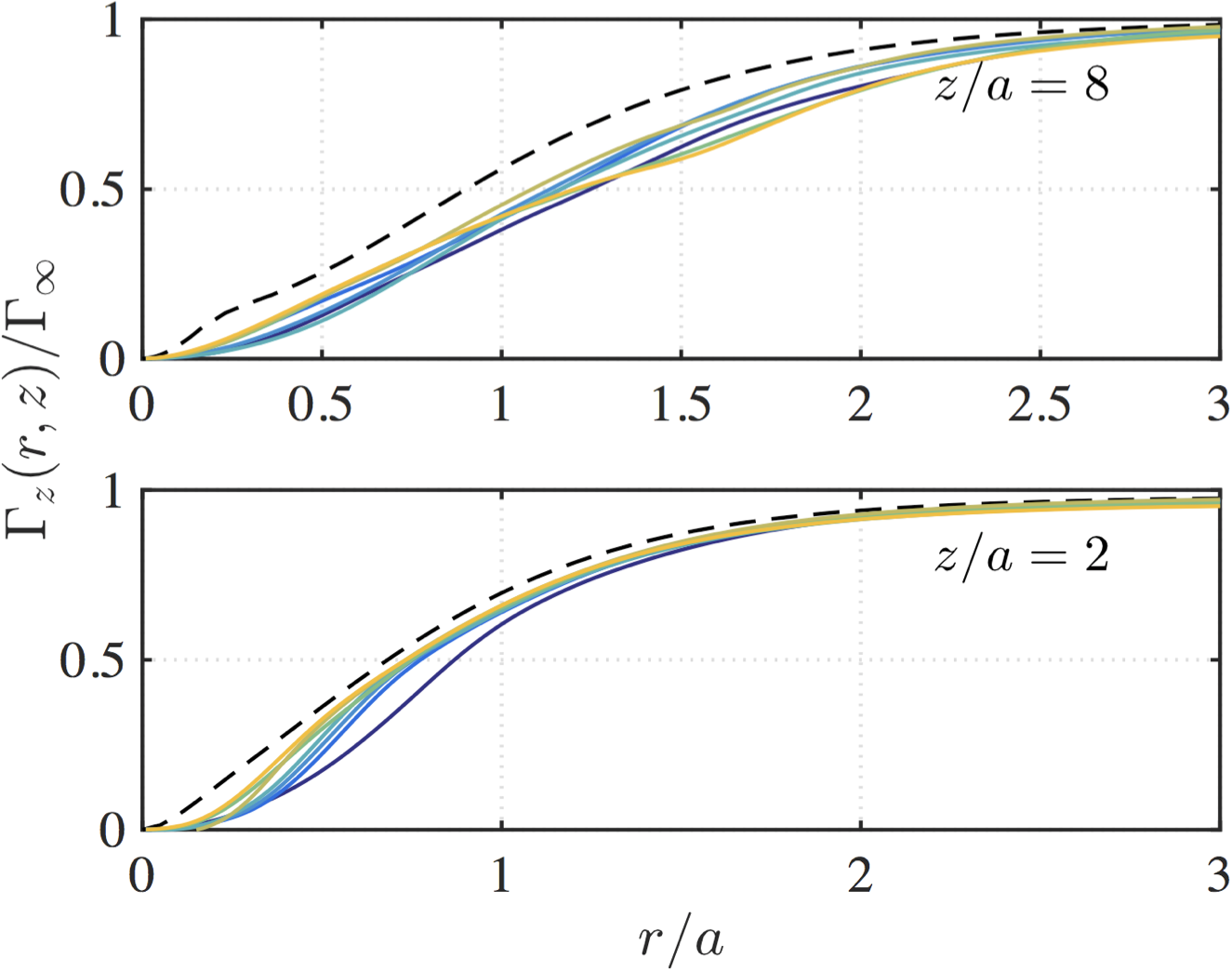} &
    \includegraphics[width=0.47\textwidth]{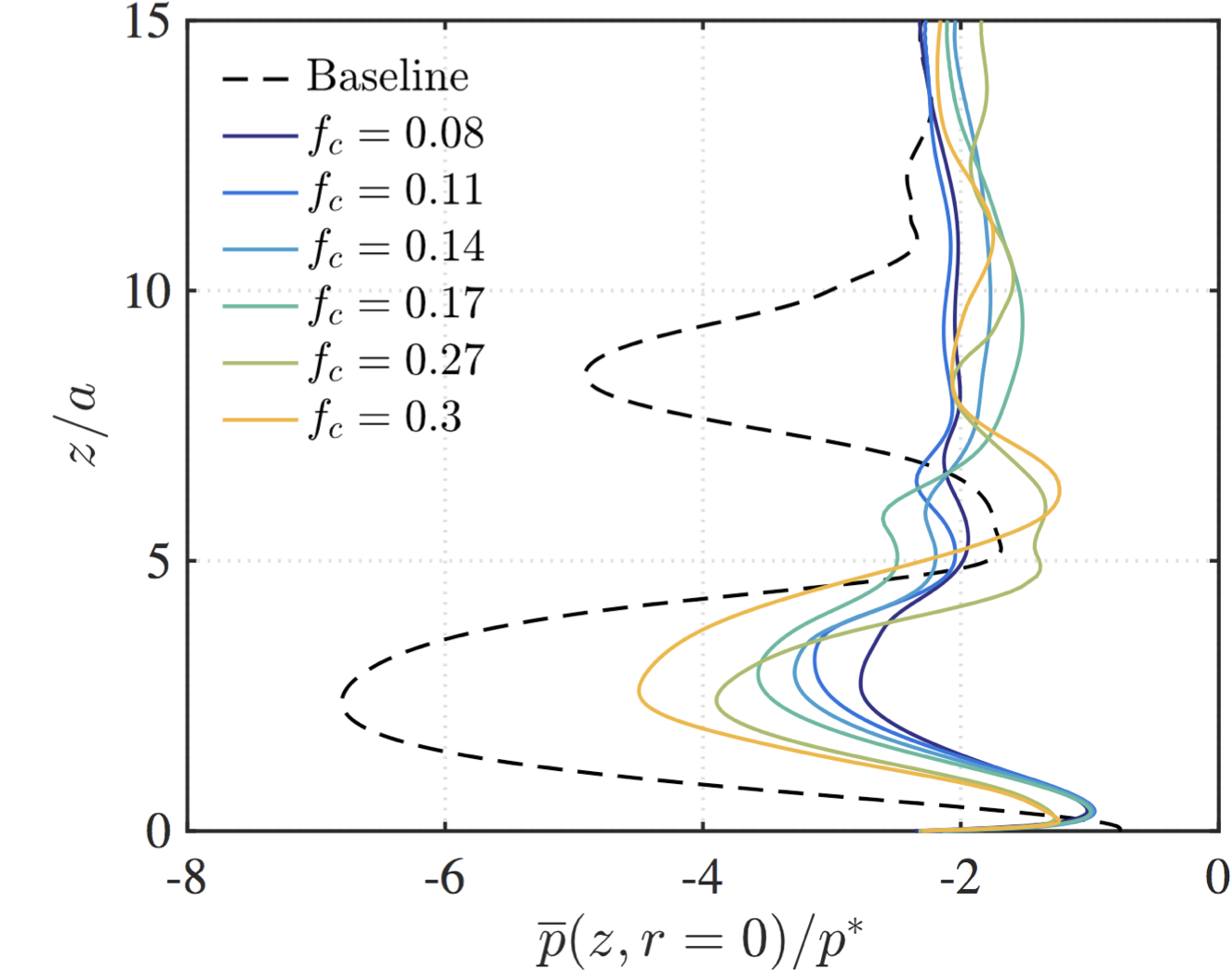} 
\end{tabular}
\caption{(a) Mean circulation of baseline and co-rotating control cases at $A=1$. (b) Time-averaged pressure distribution along the vortex center for different control frequencies.}
\label{fig:freq_effects}
\end{figure}

While keeping the actuation amplitude constant at $A_c=1$ ($C_\mu^\Gamma = 2.03\%$), let us vary the actuation frequency $f_c$ from $0.08$ to $0.3$.  We present the time-averaged axial circulation $\Gamma_z(r,z)/\Gamma_\infty$ at representative axial locations in figure \ref{fig:freq_effects}(a). With actuation, we observe that the core of the wall-normal vortex is spread over the radial direction.  The actuation introduced from the bottom wall is able to generate significant level of unsteadiness around the vortex core to eliminate the vortex breakdown region and decrease the azimuthal velocity, which in turn raises the core pressure profile, as it can be seen in figure \ref{fig:freq_effects}(b). Among the actuation frequencies considered, those near $f_c \approx 0.1$ are found to be beneficial for pressure increase.  In fact, the controlled flow with $f_c = 0.08$ achieves the highest increase in pressure for $z/a<5$.  We will further elaborate on the reason for low frequency actuation being effective in modifying the vortical flow.

\begin{figure}
\begin{tabular}{ll}
(a)  & (b) \vspace{2mm} \\
\includegraphics[width=0.47\linewidth]{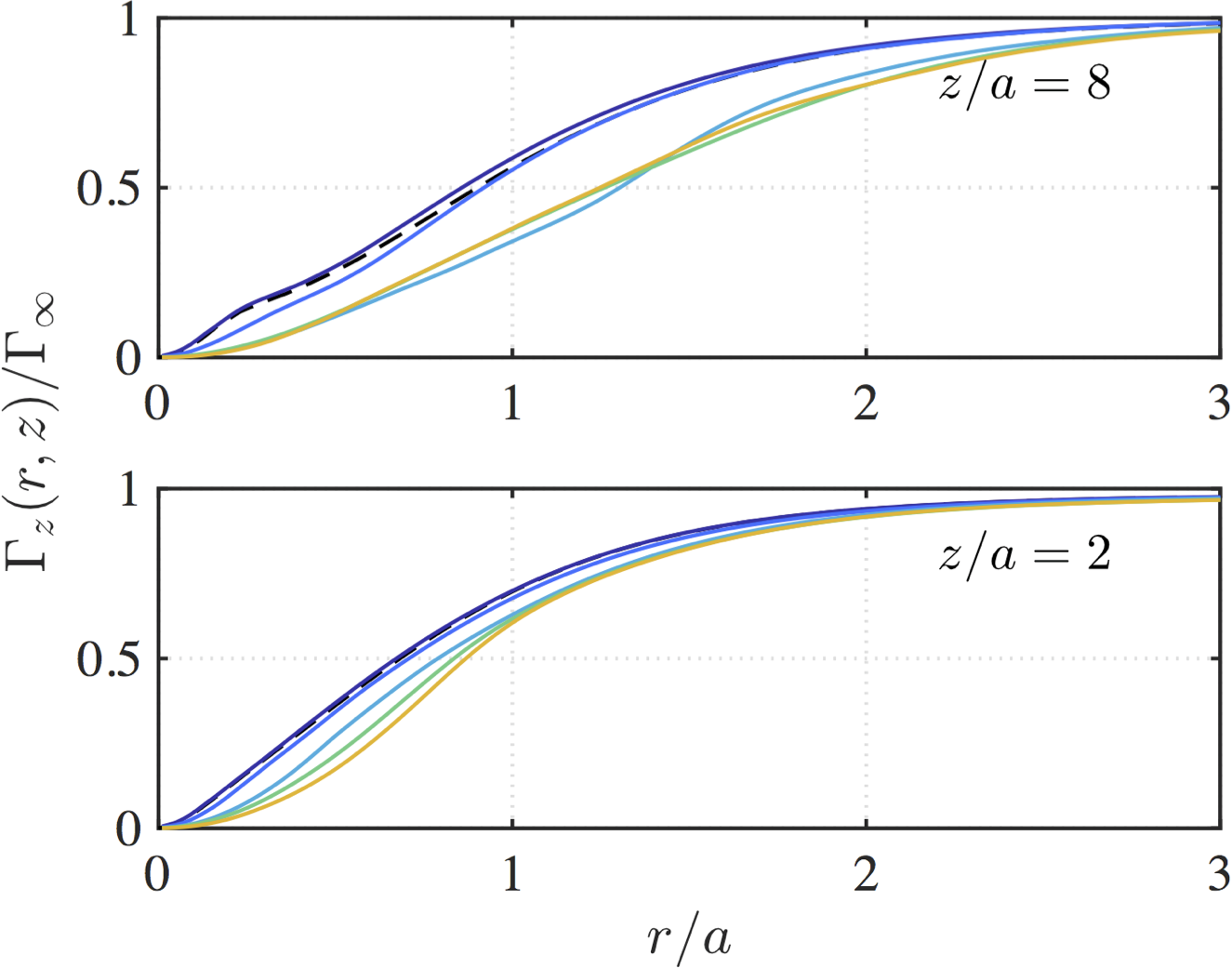} &
\includegraphics[width=0.47\linewidth]{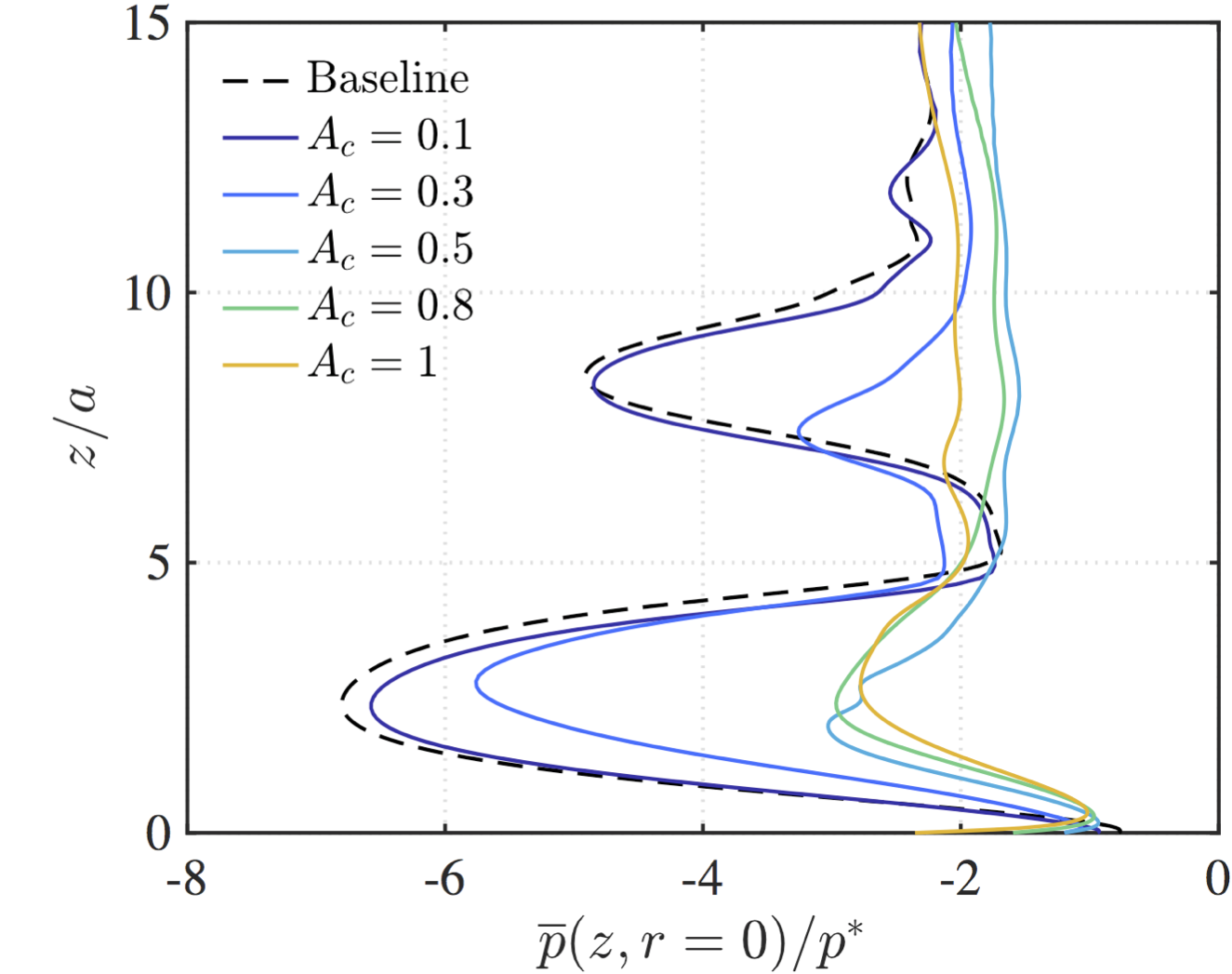}
\end{tabular}
\caption{Time-averaged (a) axial circulation at representative locations and (b) pressure distribution along the vortex core $r=0$ for varied control amplitudes (co-rotating cases) at $f_c = 0.08$.}
\label{fig:amp_effects}
\end{figure}

We further examine the impact of the control amplitude at the low frequency of $f_c=0.08$ using co-rotating actuation, which achieves significant core pressure increase. The forcing amplitude is varied from $A_c = 0.1$ to $1$, as shown in figure \ref{fig:amp_effects}. The time-averaged axial circulation $\Gamma_z(r,z)/\Gamma_\infty$ at low pressure locations of $z/a=2$ and 8 are shown in figure \ref{fig:amp_effects}(a). For $A_c\le 0.3$ $(C_\mu^\Gamma \le 0.18\%)$, the actuation does not substantially alter the circulation profile. For $A_c\ge 0.5$ $(C_\mu^\Gamma \ge 0.51\%)$, the circulation is significantly reduced and the core distribution becomes wider in the radial direction. The control becomes highly effective past $A_c=0.5$ $(C_\mu^\Gamma = 0.51\%)$. Further increasing the control amplitude, the circulation continues to decrease but the amount of reduction saturates.

The corresponding  time-averaged pressure distribution along the vortex core is presented in figure \ref{fig:amp_effects}(b).  For $A_c=0.1$ and $0.3$, the time-averaged pressure profile does not deviate much from that of the baseline. As the control amplitude increases to $A_c = 0.5$ ($C_\mu^\Gamma = 0.51\%$), the vortex breakdown region is removed and the time-averaged pressure substantially increases. This indicates the existence of a critical threshold amplitude $A_c \approx 0.5$ to achieve effective flow control for modifying the pressure distribution. Once the control amplitude surpasses $A_c > 0.5$, there is no significant gain from using the additional control input.  This tells us that the pressure profile of this type of flow can be altered effectively at the threshold amplitude using low-frequency actuation.  

We have observed that low-frequency forcing is effective in modifying the wall-normal vortex profile and increasing its core pressure.  To shed light on the influence of frequency on altering the base flow, let us revisit the Burgers vortex model.  In the three-dimensional stability analysis of the Burgers vortex performed by \cite{Schmid:JFM04}, it was reported that there is significant transient growth of perturbation that can take place (while being neutrally stable).  They performed the stability analysis by considering the Lundgren transformation \citep{lundgren1982strained} to reduce the three-dimensional flow to be described by a mapped two-dimensional diffusion equation.  In this reduced state, they quantified the gain of the transient growth $\hat{G}$ as a function of the axial wavenumber $\hat{k}_z$, which is non-dimensionalized as $\hat{k}_z = \overline{k}_z \kappa(t_\text{max})$ with $\overline{k}_z$ being the wavenumber in the mapped state and $\kappa$ being a scaling parameter that is dependent on the time of maximum energy amplification \citep[for details, see][]{Schmid:JFM04}.

\begin{figure}
\hspace{3.2cm}\includegraphics[height=0.38\linewidth]{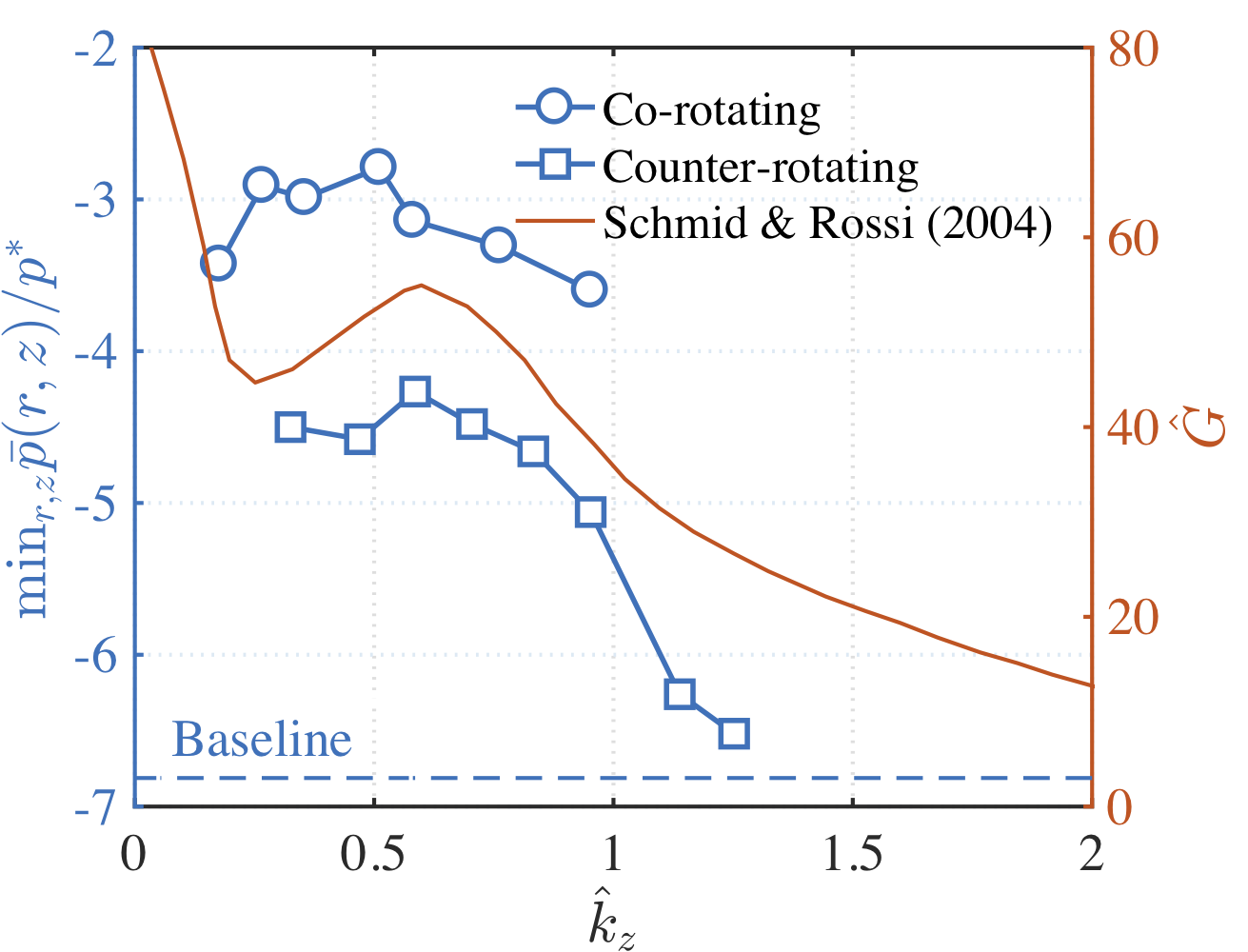}
\caption{Minimal spatial pressure shown as a function of the non-dimensional axial wavenumber $\hat{k}_z$. Also shown is the energy amplification (transient growth gain) from the stability analysis of \cite{Schmid:JFM04}.  All results are shown for $Re_\Gamma = 5000$.  The dashed line shows the minimal core pressure for the baseline flow.}
\label{fig:phys_wave}
\end{figure}

Here, we consider their stability analysis as a simplified model for the present study, as we expect some commonality in behavior especially away from the wall. The spatial minimum pressure value achieved with $A_c = 1$ for $0.02 \le  f_c \le 0.17$ (co-rotating) and $0.08 \le  f_c \le 0.3$ (counter-rotating) over the non-dimensional axial wavenumber $\hat{k}_z$ are shown in figure \ref{fig:phys_wave}. Also plotted is the transient growth gain for the Burgers vortex at $Re_\Gamma = 5000$ reported by \cite{Schmid:JFM04} to show a qualitative comparison.  We can note that sizable transient growth can be expected for $\hat{k}_z \le 1$, which corresponds to $f_c \le 0.17$ (co-rotating) and $f_c \le 0.23 $ (counter-rotating) in the present study.  The pressure changes attained by co-rotating and counter-rotating control cases are provided, which exhibit qualitative agreement with the insights on transient growth.  Hence, we infer that the transient growth mechanism (secondary hump around $\hat{k}_z=0.7$) plays an important role in ensuring that small forcing inputs are greatly amplified to effectively modify the mean pressure profile of the wall-normal vortex.  Because co-rotating flow control creates large unsteadiness in an nonlinear manner, it is expected that there is some difference in wavenumber that achieves the highest mean flow modification and transient growth gain distribution.  The counter-rotating case on the other hand agrees well with the transient gain distribution, while the perturbation does not grow as much as the co-rotating control cases, as we can observe from figure \ref{fig:vortex_strucuture}.  Since the forcing is harmonic in time, resolvent analysis \citep{jovanovic2005componentwise, mckeon2010critical, yeh2018resolvent} would also be required to gain a deeper understanding of the control mechanism. The insights from non-modal analysis can offer guidance on finding an effective control setup to modify the mean flow and reducing the control parameters space to search over.

At last, we note that cases with $m=0$ and $|m|>1$, as well as steady control with $f_c=0$ were also considered.  We have found that cases with $m=0$ and $|m|>1$ do not modify the vortical flow as effectively as the cases with $m=\pm 1$.  The observation of the control setup with $m=1$ working most effectively agrees qualitatively with the non-modal analysis of Burgers vortex conducted by \cite{Schmid:JFM04}. With the steady actuation $f_c=0$, the vortex core displaces away from $r=0$ and still exhibits axial variation (i.e., $\hat{k}_z \neq 0$).   The results imply that unsteady actuation plays an important role in effectively alleviating the low vortex core pressure.

\section{Concluding Remarks}
\label{sec:conclusion}

We have characterized and controlled a wall-normal vortex based on the Burgers vortex inflow condition in direct numerical simulation at a circulation based Reynolds number of 5000.  This type of vortex often appears upstream of turbomachinaries and can cause degradation in operating efficiency and potential structural damage.  As such, the objective of the present study was to increase the pressure of the vortex core and avoid the potential appearance of hollow-core vortices by modifying the vortical flow through active flow control. 

The baseline vortical flow was found to have a steady near-wall region, a vortex breakdown region, and a wake region.  Dynamic mode decomposition was performed on the flow field data to identify the unsteady modal structures and their frequencies.  Based on the insights gained from baseline flow characterizations, we have developed an active flow control approach where unsteady blowing and suction (zero net mass flux) is introduced from the bottom wall at the center of the vortex.  The forcing input was added in an unsteady manner with co-rotating and counter-rotating setups.  We have observed that co-rotating control is more effective in altering the vortical flow.  The control input is able to spread the vortex core and increase the level of unsteadiness, which eliminates the vortex breakdown region.  With the core widening radially, the azimuthal velocity decreases and the core pressure increases.

Moreover, the influence of actuation frequency and amplitude was analyzed.  For co-rotating control, it was observed that low-frequency forcing is effective in weakening the core region of the vortex, providing significant pressure increase. We have also found that there is a threshold forcing amplitude (circulation based momentum coefficient of $C_\mu^\Gamma = 0.51\%$), above which the control effect essentially saturates. The developed control setup is found to take advantage of the transient growth of perturbation input and force the mean flow to be altered.  Such observation qualitatively agrees with non-modal stability analysis of Burgers vortex reported by \cite{Schmid:JFM04}.  The insights from non-modal analysis has offered guidance on finding an effective control approach to modify the mean flow and reduce the space over which to search for appropriate control parameters.  The present flow control technique opens a promising path to alleviate adverse effects from strong wall-normal vortices. 


\bibliographystyle{jfm}
\bibliography{Taira_refs,references,references2}
\end{document}